# An Open-Access Database of Active-source and Passive-wavefield DAS and Nodal Station Measurements at the Newberry Florida Site

Aser Abbas [a, *], Brady R. Cox [a], Khiem T. Tran [b,] Isabella Corey [a], Nishkarsha Dawadi [a]

[a] Utah State University, Department of Civil and Environmental Engineering, Logan, UT, USA, 84322.
[b] University of Florida, Department of Civil and Coastal Engineering, 365 Weil Hall, P.O. Box 116580, Gainesville, FL 32611, USA.

**Abstract**

This paper documents a comprehensive subsurface imaging experiment using stress waves in Newberry, Florida, at a site known for significant spatial variability, karstic voids, and underground anomalies. The experiment utilized advanced sensing technologies, including approximately two kilometers of distributed acoustic sensing (DAS) fiber optic cable, forming a dense 2D array of 1920 channels, and a 2D array of 144 three-component nodal stations, to sense active-source and passive-wavefield stress waves. The active-source data was generated using a vibroseis shaker truck and impact sources, and it was simultaneously sensed by both the DAS and the nodal stations. The vibroseis truck was used to excite the ground in the three directions at 260 locations inside and outside the instrumented array, while the impact sources were used at 268 locations within the instrumented array. The passive-wavefield data recorded using the nodal stations comprised 48 hours of ambient noise collected over a period of four days in four twelve-hour time blocks. Meanwhile, the passive wavefield data collected using DAS consisted of four hours of ambient noise recordings. This paper aims to provide a comprehensive overview of the testing site, experiment layout, the DAS and nodal station acquisition parameters, implemented processing steps, and potential use cases of the dataset. While potential use cases, such as surface wave testing, full waveform inversion, and ambient noise tomography, are discussed relative to example data, the focus of this paper is on documenting this unique dataset rather than on processing the data for detecting anomalies or generating subsurface 2D/3D imaging results. The raw and processed data, along with detailed documentation of the experiment and Python tools to aid in visualizing the DAS dataset have been archived and made publicly available on DesignSafe under project PRJ-3521.

*Corresponding author.
E-mail: aser.abbas@usu.edu (A. Abbas)

**Introduction**

Noninvasive imaging techniques are increasingly being used for geotechnical site characterization due to their advantages in time, cost, and spatial coverage when compared to traditional invasive methods. Geophysical imaging based on stress wave propagation continues to evolve, with new innovations emerging to meet increasingly complex demands, such as higher imaging resolution for elastic moduli, anomaly detection, and damping estimation. High-quality field data is essential for developing and testing these emerging techniques. This paper presents a comprehensive and open-access dataset of stress wave recordings gathered using some of the most advanced technologies available in geophysical-noninvasive subsurface imaging. A test site in Newberry, Florida was selected for this extensive subsurface imaging experiment due to its complex geology, which includes many known and unknown karstic voids of variable size and depth. A 2D layout of distributed acoustic sensing (DAS) fiber optic cable and a 2D array of three-component (3C) geophone nodal stations covering an area approximately 155 m x 75 m were used at the site to record both active-source and passive-wavefield stress waves (refer to Figure 1). The active sources used to initiate seismic wave propagation comprised both a broadband, three-dimensional, vibroseis shaker truck named T-Rex from the NHERI@UTexas experimental facility (Stokoe et al., 2020), and more-variable, narrow-band, impact sources. In total, approximately 2 km of DAS fiber optic cable and 144, 3C nodal stations were used to record wavefields from more than 367 shot locations. This unique and publicly accessible dataset is available on DesignSafe (Rathje et al., 2017; https://www.designsafe-ci.org/) under project PRJ-3521, "Active-source and Passive-wavefield DAS and Nodal Station Measurements at the Newberry Florida Site". The ensuing paragraphs offer a concise overview of the sensing technologies employed in this experiment and the potential value of the dataset documented herein.

The first sensing technology used in this experiment was DAS, which is a rapidly evolving technique for transforming fiber-optic cables into a distributed array of ground motion sensors (Cox et al., 2012; Yu et al., 2019). It is increasingly being used to sense active and passive stress waves for geophysical imaging and seismic monitoring of the near surface (e.g., Dou et al., 2017; Hubbard et al., 2022; Vantassel et al., 2022). DAS measures dynamic strain by using an interrogator unit (IU) to fire a series of laser pulses (probe pulses) through a fiber optic cable. The interaction between a probe pulse and the fiber's inhomogeneities returns a backscattered signal, mainly composed of Rayleigh backscatter, to the launching end. The relative phase of the Rayleigh backscattered light is used to determine changes in length between scattering regions along the cable. This determination is repeated for each resolvable point along the fiber, resulting in a measure of the dynamic strain as a function of time and location (Hartog, 2018). Studies by Daley et al. (2016), Hubbard et al. (2022), and Vantassel et al. (2022) have demonstrated that DAS measurements can yield very similar waveforms and processed data (e.g., surface wave dispersion data) as geophones when proper care is taken. Furthermore, DAS can provide unprecedented spatial resolutions (on the order of meters) and length scales (on the order of tens of kilometers), surpassing traditional sensing technologies (Soga & Luo, 2018). For instance, DAS measurements with the ~2-km long fiber optic cable used in the present study resulted in 1920 channels of vibration data sensed at a spatial resolution of 1.02 m, which is equivalent to deploying 1920 unidirectional geophones. In recent years, several studies have been conducted to evaluate the potential of using DAS for non-invasive near-surface imaging. Most of these studies utilized either a 1D (i.e., line) DAS cable (e.g., Hubbard et al., 2022; Vantassel et al., 2022), a 2D DAS cable configuration and an impact source (e.g., Castongia et al., 2017), or a 2D DAS cable configuration and ambient noise (e.g., Dou et al., 2017). Lancelle et al. (2014) employed a 2D DAS fiber optic cable configuration along with shear and vertical vibrational sources at the Garner Valley testing site in California, United States (US). However, the cable runs were sparsely spaced and the sources were utilized at a limited number of locations. Obermann et al. (2022) conducted a seismic study in the Hengill geothermal area in southwest Iceland using a network of 3C nodal stations and two DAS fiber optic cables, along with a vibroseis shaker. Their research focused on imaging the top four kilometers of the crust, and thus, their nodal stations were spaced out over several kilometers in each direction, with interstation distances varying from tens to hundreds of meters. Furthermore, their DAS cables had a nearly linear configuration, and the vibroseis source only generated vertical vibrations. In contrast, the experiment documented herein focuses on the near surface depths relevant to geotechnical engineering (less than ~ 50 m). The study stands out for its use of a densely spaced 2D configuration of DAS cable to capture both passive- and active-wavefields generated by vibroseis shaking in three directions. These wavefields were simultaneously recorded by a dense 2D array of 3C geophone nodal stations, allowing for a detailed comparison of DAS and nodal station measurements, further setting this dataset apart from prior work.

The second sensing technology deployed at the test site was 3C nodal stations, which allow for the concurrent measurement of ground shaking in all three directions. This complements the unidirectional sensitivity of the DAS system, thereby providing a more comprehensive view of the seismic wavefield, albeit at approximately 1/5 the spatial resolution of the DAS measurements across the site. The three perpendicular geophones (two horizontal and one vertical) exhibit sensitivity to different types of waves. For instance, Rayleigh and compression waves are best identified on the vertical and horizontal inline components, while the horizontal crossline component is better suited to detect Love waves (Foti et al., 2018; Vantassel, Cox, et al., 2022; Vantassel & Cox, 2022). Since different wave types carry information about different mechanical properties of the subsurface (Sheriff & Geldart, 1995), analyzing the data collected by 3C nodal stations can provide valuable insights into the 3D mechanical properties of the site. The versatility of the 3C nodal station measurements open avenues for researchers to use the dataset who are working on techniques that leverage either the vertical, any of the horizontal, or any combination of the three directions of ground shaking measurements (e.g., Cheng et al., 2020, 2021; Cox et al., 2020; Fathi et al., 2016; Kristekova et al., 2020; Nakamura, 1989; Pan et al., 2016; Smith et al., 2019; Wang et al., 2019; Wathelet et al., 2018).

The karstic voids at the testing site add an intriguing dimension to this experiment. Several noninvasive seismic techniques have been developed to detect voids and other underground anomalies (e.g., Belfer et al., 1998; Branham & Steeples, 1988; Cook, 1965; Kolesnikov & Fedin, 2018; Kristekova et al., 2020; Pernod P. and Piwakowski, 1989; Sloan et al., 2012; Smith et al., 2019; Wang et al., 2019). However, detecting anomalies using real field data remains challenging (Grandjean & Leparoux, 2004; Sloan et al., 2010; Smith et al., 2019). For example, Smith et al. (2019) and Wang et al. (2019) conducted an experiment to image a known 0.9 m x 1.5 m x 96 m tunnel situated ten meters below the surface at Yuma Proving Ground, Arizona, US. The experiment utilized a dense 2D array comprising 720 vertical and 720 horizontal geophones and an accelerated weight-drop source. They used both 2D and 3D full waveform inversion (FWI) to image the tunnel. Smith et al. (2019) highlighted the advantages of using various source orientations with multicomponent seismic sensors when imaging for voids. Their research revealed that different combinations of source orientations with receiver components produced varying resolutions of the tunnel. The Smith et al. (2019) and Wang et al. (2019) studies also showed that 2D FWI was effective in imaging the known location of the tunnel, because the experiment was designed specifically for that purpose. However, for imaging unknown void locations with complex shapes, 3D FWI would be more suitable. Smith et al. (2019) noted that the resolution of the tunnel was limited by the lack of higher-frequency data used in the inversion. We anticipate that the present dataset, featuring a powerful, triaxial vibroseis shaker and 3C sensors, in conjunction with the dense DAS array, will serve as a valuable resource for researchers seeking to explore novel approaches for void imaging.

**Site Overview**

The test site is a dry retention pond located in Newberry, Florida (29°39'0.39" N, 82°35'51.20" W) along State Road 26 (refer to Figure 1). Sinkholes are common in this area and generally an immense problem in parts of Florida due to karstic geology. The Newberry retention pond site has undergone thorough investigations over the years, utilizing both invasive and non-invasive methods. Findings from these previous studies indicate that the subsurface is comprised of medium-dense fine sand and silt that range in depth from two to ten meters overlying highly variable limestone (Tran et al., 2013; Tran & Hiltunen, 2011). The site also contains sinkholes of different sizes and depths, some of which have been repaired. The varied stiffness and depth of the limestone layer, along with the presence of surficial and underground anomalies, make this site a prime location for noninvasive subsurface imaging research. Tran and Hiltunen (2011) conducted ten cone penetration tests (CPT), eight geotechnical borings with standard penetration tests (SPT), and 12 consecutive seismic refraction tests using a linear array of 31, 4.5-Hz vertical geophones and a sledgehammer source. The first arrival times from the refraction tests were inverted using simulated annealing to develop 2D compression wave velocity ($V_P$) profiles of the site. However, no voids were identified via seismic refraction testing by Tran and Hiltunen (2011). Tran et al. (2013) used a linear array of 24, 4.5-Hz vertical geophones and a sledgehammer source to collect seismic data at ten different locations at the site. The seismic data was then inverted using 2D FWI, which identified an underground anomaly that was later verified to be a void through an SPT sounding. Nonetheless, they observed that the predicted depth of the void was greater than its actual depth. They attributed this discrepancy to the difference between the measured wavefield, which is affected by the three-dimensional variations in the subsurface, and the assumed plain strain condition used in their 2D FWI. Tran et al. (2020) expanded on the FWI studies at the site by utilizing the one-dimensional, small vibroseis shaker truck named Thumper from the NHERI@UTexas experimental facility (Stokoe et al., 2020) to excite the ground at 65 locations within and around a 2D grid of 48, 4.5-Hz vertical geophones arranged in a 4 x 12 configuration. The source and receiver 2D grids were uniform, with 3-m spacing, covering a total area of 12 m x 36 m. Using the collected data and 3D FWI analyses, Tran et al. (2020) created a 3D subsurface model below the sensor grid, identifying a low velocity anomaly and a void that was confirmed through an SPT sounding.

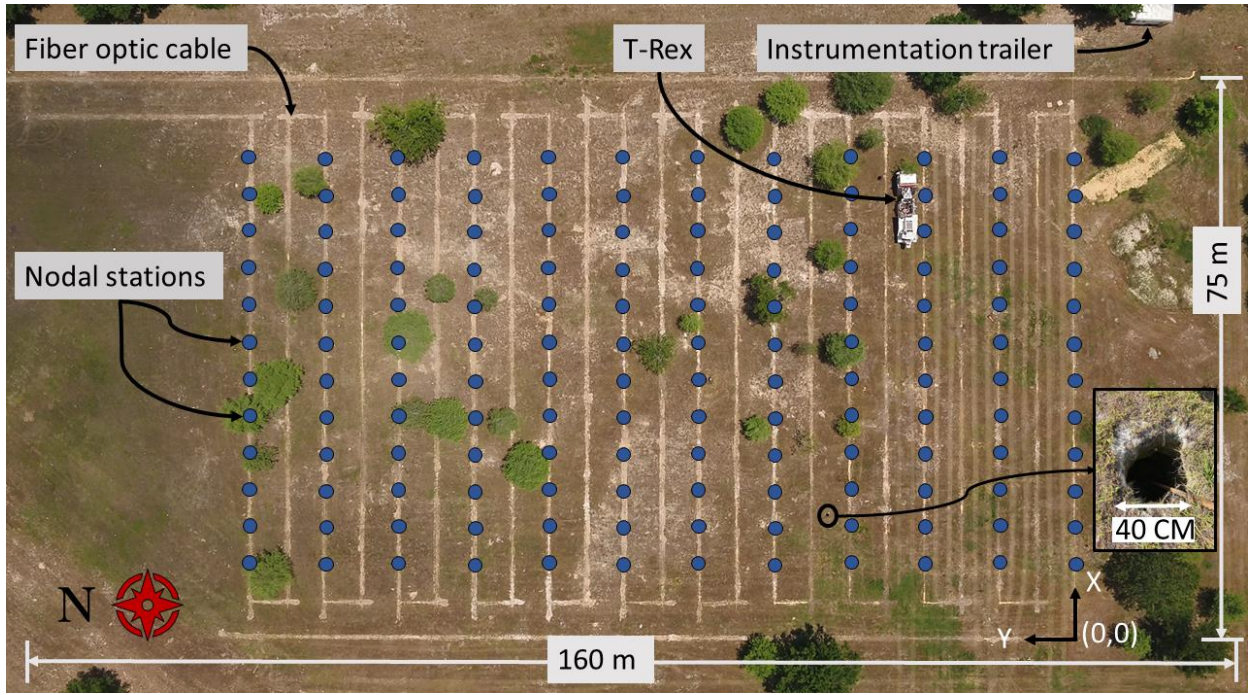

**Figure 1.** Overhead-view of the Newberry test site captured from a drone, showing the trenches used for fiber optic cable installation, as indicated by white, linear markings on the ground surface, and the 3C geophone nodal station locations, indicated by blue circle symbols. Additionally, the figure highlights one of the T-Rex shot locations, one of the voids present inside of the array, and the instrumentation trailer outside of the array.

Despite being used for a number of previous subsurface imaging studies, the authors felt there was an opportunity to collect a unique subsurface imaging dataset at the Newberry site that would improve upon previous studies by: (a) covering a larger spatial area, (b) using DAS to enable much more dense spatial sampling, (c) using 3C nodal stations to enable multi-component processing, (d) using a dense grid of more numerous shot locations, and (e) incorporating 3C shaking capabilities from a broadband and powerful vibroseis shaker truck. It is hoped that this will enable more advanced processing techniques to be applied, both in the present and in the future, with the goal of achieving deeper and higher resolution imaging to resolve subsurface anomalies.

**Overview of the Dataset**

The subsurface imaging dataset documented in this paper was collected over the course of eight days, beginning Monday, May 9th, 2022, and ending Monday, May 16th, 2022. The dataset consists of both active-source and passive-wavefield stress wave recordings that were sensed using 2D arrays of DAS fiber optic cable and 3C nodal stations. Approximately two kilometers of continuous DAS fiber optic cable was laid out in the zigzag pattern shown schematically by a black line in Figure 2, which is also visible through the white markings from the fiber optic cable trenches shown in the drone image presented in Figure 1. The fiber optic cable was interrogated using an OptaSense ODH 4+ interrogator unit (IU) configured to have a 1.02-m channel separation, resulting in a total of 1920 channels. A 2D array of 144, 3C geophone nodal stations was also deployed on site, as indicated by the blue-solid circles in Figures 1 and 2. These stations were arranged in a 12 x 12 grid, evenly spaced every five and ten meters in the X (west-east) and Y (south-north) directions, respectively. The active-source data was generated from 260 shot locations inside and outside of the instrumented area where T-Rex was used to shake the ground in all three directions, and 286 shot locations inside the instrumented area where impact sources were used to strike the ground vertically (refer to Figure 2). The passive wavefield data consisted of approximately four hours of ambient noise recordings using DAS, and approximately 48 hours of ambient noise recordings using the nodal stations. The dataset is permanently archived and publicly available on DesignSafe (Rathje et al., 2017; https://www.designsafe-ci.org/) under a project titled "Active-source and Passive-wavefield DAS and Nodal Station Measurements at the Newberry Florida Site". The project houses three parent folders: a "Raw data" folder, which contains both the DAS and the nodal station data in their original, as-collected form; a "Processed data" folder, which organizes the data in a user-friendly format after undergoing preprocessing; and a "Supporting documents" folder, which includes complete and thorough documentation of the experiment. The experimental setup and dataset organization are explained in detail in the following sections.

**Experiment layout**

The experiment was laid out on a 2D survey grid, as illustrated in Figure 2. The grid consisted of 25 horizontal lines pointing approximately east (bearing 89°) and 16 vertical lines pointing approximately north (bearing 359°). The horizontal lines, except for the lowermost and uppermost, were uniformly spaced at five meters and were labeled with the letters A through W in order from south to north, respectively. The lowermost line, Z, was 15 m south of line A, while the uppermost line, ZZ, was 30 m north of line W. The vertical lines were spaced five meters apart and named from west to east, 101, 102, 1, 2, 3, … through 12, 103, and 104, respectively. The grid intersection points will be referenced first by the letter and then by the number representing the intersecting horizontal and vertical lines, respectively (e.g., A101). Although the global latitude and longitude coordinates of the grid points were surveyed and are included in the electronic dataset, this paper will utilize a local coordinate system for ease of reference, with the origin at point A101 (local coordinate 0,0), the positive X direction pointing eastward, and the positive Y direction pointing northward.

The site is relatively flat and the grass was generally quite short at the time of testing. The lines and points of the survey grid, which mark the fiber optic cable and nodal station locations, as detailed in subsequent sections, were surveyed into position using two total stations and several 100-m tape measures. The first total station was deployed at point A102 (coordinate 5,0) and was oriented in the positive direction of the Y axis (refer to Figure 2). With the bearing set, the intersection points between all the horizontal lines (e.g., B, C, D, etc.) with line 102 were established. The total station was then rotated 90° towards the east to align with the positive direction of the X axis. With this bearing set, the intersection points between all the vertical lines (e.g., 1, 2, 3, etc.) with line A were established. At that point, the second total station was deployed at point A103 (coordinate 70,0) to survey the intersection points between all the horizontal lines (e.g., B, C, D, etc.) with line 103. This process was repeated for all lines of the grid, with forward- and back-sights established whenever a total station was relocated to maintain orthogonality. Projecting the experiment layout onto the test site was completed in the first day, although minor surveying was performed throughout the experiment to ensure the proper placement of the fiber optic cable, nodal stations, shots, etc.

**Distributed Acoustic Sensing (DAS)**

Selecting the right fiber optic cable is crucial for good DAS measurements, since the cable functions as both the strain sensing element and the means of transmitting optical signals (Hartog, 2018). In this experiment, a fiber optic tactical cable (AFL X3004955180H-RD) consisting of four tight buffered fibers coated in a layer of aramid yarn and enclosed by a polyurethane jacket was used. Studies by Hubbard et al. (2022) and Vantassel et al. (2022) confirm that this cable offers good deformation coupling between the internal optical fiber and the ground when buried with soil compacted around/over it. The length of cable installed at the site was approximately 2 km (exactly 1958.4 m), with one end connected to the IU at the instrumentation trailer located in the southeast corner of the testing site (refer to Figure 1), and the other end appropriately terminated at the northwest corner of the site to reduce end reflections. The cable was buried in a shallow trench to ensure optimal coupling between the ground and the cable. To facilitate precise trenching along the grid lines, a total station was positioned at the endpoint of each cable route and a tape measure was pulled tight to mark the trenching path. A second tape measure was then horizontally offset from the first by an amount equal to the distance between the trencher blade and its tire, as illustrated in Figure 3a. The trencher was guided along the surveyed line to create a trench that was approximately 20-cm wide and 10- to 15-cm deep, ensuring its straightness. The cable was then manually placed at the bottom of the trench by rolling the cable spool over it (Figure 3b), with slight tension applied to minimize slack. The corners of the trench were rounded to a radius of approximately 20 cm (Figure 3c), which is greater than the AFL cable's allowable bend radius. Following cable installation, the trench was backfilled with native soil, or with clean sand when the native soil was too hard and clotted to allow for good coupling (Figure 3d and 3e). The backfilled soil was then manually compacted over the cable to ensure good coupling with the native ground (Figure 3f). All cable corners were left exposed until tap tests could be performed to index the DAS cable (i.e., map the DAS channel numbers to their physical locations). The tap tests involved lightly tapping on the fiber optic cable at all corners and other important locations (such as the start and end of the cable) and noting the DAS channels that responded with significant energy. Based on the tap tests, the first and last channels on the cable with usable data (i.e., the first and last buried channels) are channels 31 and 1905, respectively, as shown in Figure 2. A tap test can only locate the measurement point with an accuracy of the gauge length of the DAS system, meaning it has a +/- one-half gauge length margin of error. As noted below with respect to DAS data acquisition, the gauge length used throughout the study was 2.04 m, so the absolute spatial location of each channel is accurate to within about 1 m. However, the relative location/spacing between channels is exactly equal to one-half of the gauge length, or 1.02 m. The corners were backfilled and compacted to ensure proper coupling between the cable and the surrounding soil. Trenching started on Monday, May 9th, from point Z104 and ended at point J102. On Tuesday, May 10th, trenching was completed, and the cable installation began from Z104, reaching J07 by the end of the day. The cable installation was completed on Wednesday, May 11th.

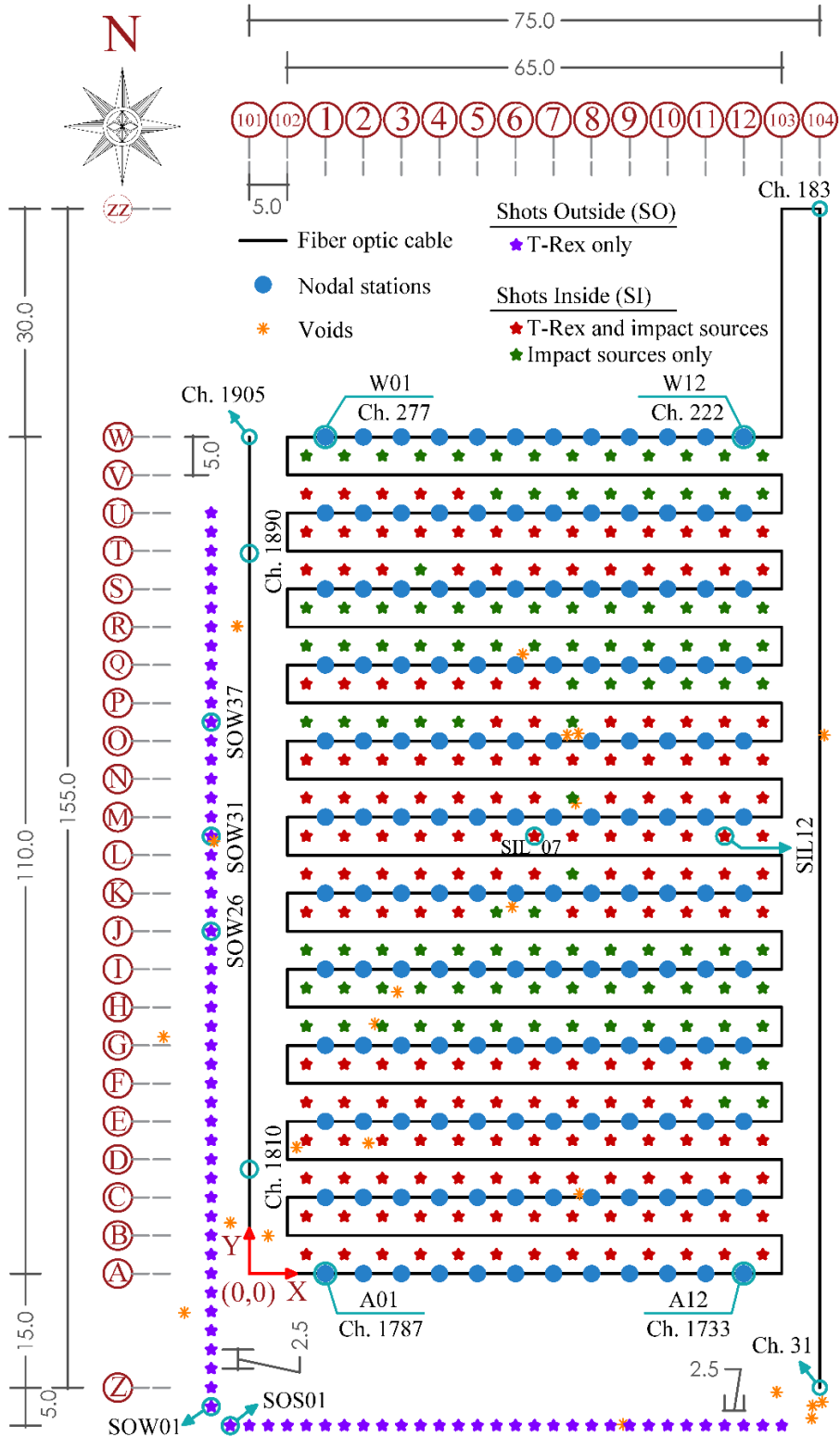

**Figure 2.** Schematic layout of the test site showing locations of the: 3C geophone nodal stations, fiber optic cable, T-Rex and impact shots, and voids that are visible from the ground surface. The layout is comprehensive, including all of the line numbers/letters and dimensions used to arrange the equipment.

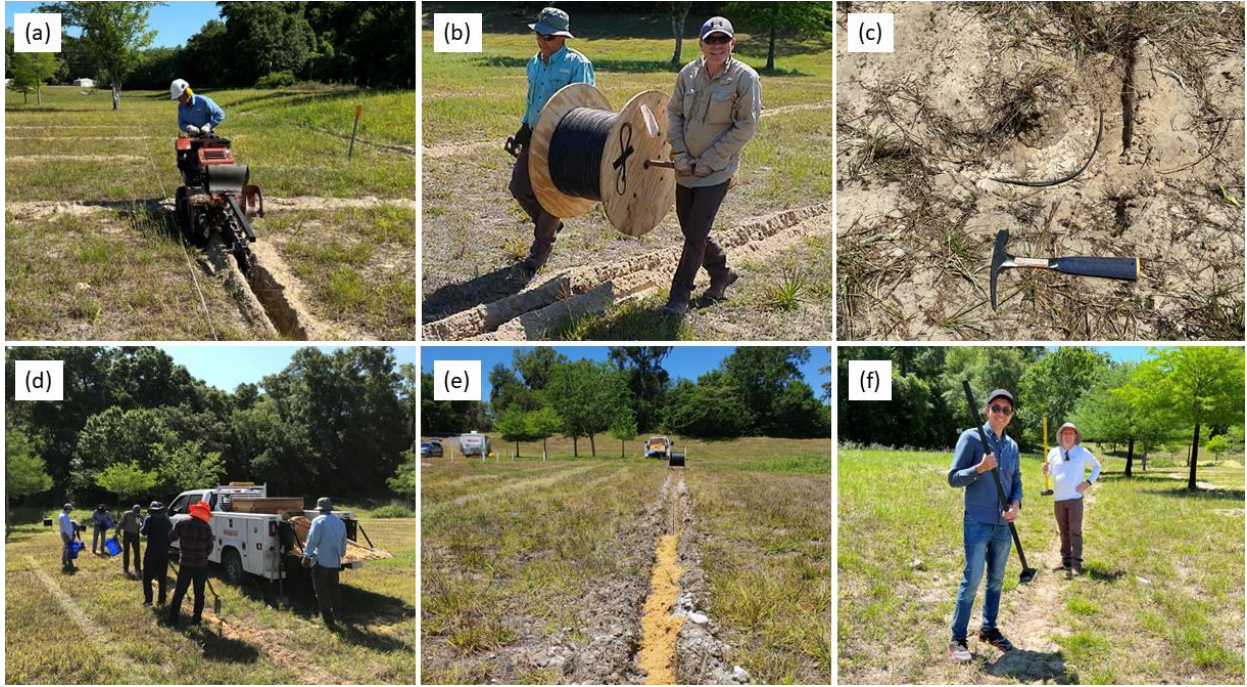

**Figure 3.** Pictures illustrating the fiber optic cable installation procedure, starting with: (a) trenching the cable route, (b) laying the cable by rolling it off the cable spool along the trench, (c) rounding the cable at the corners, (d) and (e) filling the trench with native soil or clean sand as required, and finally (f) compacting the backfilled soil to ensure proper coupling between the cable and the ground.

## Nodal stations

### Instrumentation

The nodal stations used in this experiment were SmartSolo IGU-16HR 3C. They have a compact, all-in-one modular design, with a GPS-synced, 32-bit digitizer (accurate to ±10 microseconds), a maximum input signal of ±2.5 Volts at 0 dB gain, and a storage capacity of 64 GB. Each station is equipped with three, orthogonal, 5-Hz geophones and a self-contained power supply with a 30-day battery life. The geophones are wired such that a tap from the north, east, or top causes an upward voltage departure in the geophone oriented along that axis (refer to Figure 4d). Four conical spikes were mounted on each nodal station to ensure good coupling with the ground. These IGU-16HR 3C stations have a small footprint of 95 mm x 103 mm, a height of 30 cm, and weigh around 2.4 kg with spikes attached. The 144 nodal stations used in this experiment were sourced from two locations: 88 stations from the Earthquake Engineering and Subsurface Imaging Lab at Utah State University (labeled USU01 through USU90, excluding stations USU07 and USU42), and 56 stations from the NHERI@UTexas experimental facility (labeled UT01 through UT56). The stations from both sources were independently labeled in ascending order based on their serial numbers.

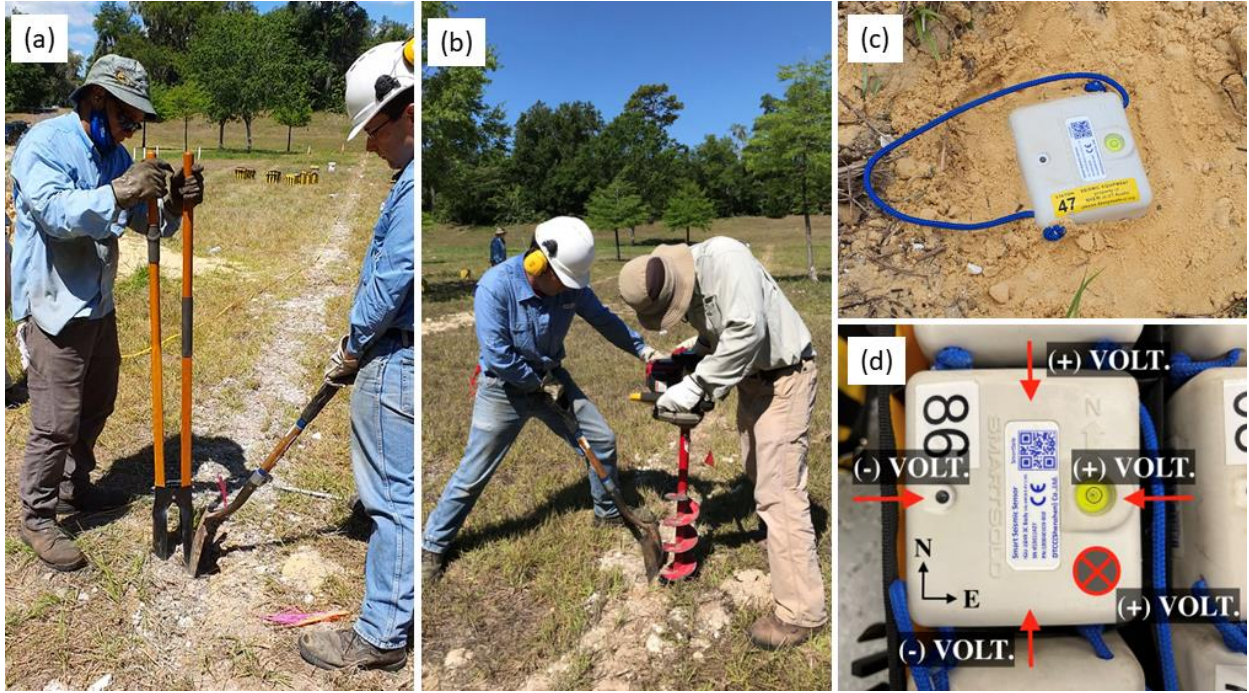

**Figure 4.** Picture illustrating the installation of nodal stations next to the fiber optic calbe. In panels (a) and (b), excavation is performed with either a post-hole digger or a gas-powered auger, respectively, while the fiber optic cable is protected with a shovel. Panel (c) depicts a completed installation, with the station securely in place and its top exposed. Panel (d) describes the voltage polarity of the three geophones in each nodal station.

*Installation*

The 144, 3C geophone nodal stations were arranged in a 12 x 12 grid configuration with ten- and five-meter spacings in the south-north and west-east directions, respectively, as shown in Figures 1 and 2. Starting with the USU stations, the stations were placed in order from south to north along line 1 (USU01 to USU13), then along line 2 (USU14 to USU25), and subsequently all the way to line 12. Remember, stations USU07 and USU42 were not used in this experiment. Thus, the UT stations started at I08 and followed the same pattern until the last station was installed at W12. The stations were deployed right next to the fiber optic cable in their respective locations. Depending on the stiffness of the ground, either a gas-powered earth auger or a manual post-hole digger was used to excavate a hole approximately 20-cm in diameter and 25-cm deep for the stations. The fiber optic cable was protected from the hole digging operations using a shovel, as shown in Figures 4a and 4b. The orientation of the stations was such that the north arrow pointed towards the +Y direction, which was orthogonal to the fiber optic cable, conveniently aligning with true north (within 1° tolerance). This ensured that the east-west geophone/channel in the three-component nodal stations was properly aligned with the fiber optic cable. The stations were then leveled and the holes backfilled using either excavated soil or clean sand when needed, leaving only the station's top exposed, as shown in Figure 4c. Following deployment of all stations, the stations were activated for continuous recording. This process involved booting up each station using a magnet switch, confirming successful activation by observing a flashing green LED state indicator, conducting a quality scan using a designated handheld device, and surveying its final location. Deploying and activating the stations started on Wednesday, May 11th, and ended on Thursday, May 12th.

**Active-source wavefield generation**

On Friday, May 13th, following installation of the fiber optic cable and all SmartSolo nodal stations, active-source wavefield recording commenced. Two source types were used for active wavefield generation: a highly controlled, powerful, broadband vibroseis source, and more variable, narrow-band, impact sources. The vibroseis source used in this experiment was the NHERI@UTexas, large mobile shaker truck named T-Rex (Figure 5a). T-Rex is a 29-ton, tri-axial vibroseis truck capable of shaking its baseplate in the vertical, longitudinal, and transverse directions (Stokoe et al., 2020). It has a maximum force output of about 267 kN in the vertical direction and 134 kN in each horizontal direction. In this experiment, T-Rex was used to generate a 12-second-long chirp with a linear frequency sweep from 5 to 80 Hz. In addition to GPS time and coordinates for each shot location, the T-Rex electronics recorded the baseplate and mass accelerations and the ground force with 1 kHz sampling rate for each shot. Herein, a "shot" refers to an instance where a source was used to excite the ground at a given location. In total, T-Rex vibrated at 260 distinct locations; 81 outside and 179 inside the instrumented area, as illustrated in Figure 2. Shots outside (SO) of the instrumented area were distributed among 30 locations to the south (S), 48 locations to the west (W), and three locations to the north (N). These locations will be referred to as SOS, SOW, and SON, respectively. Note that the three SON locations are not shown in Figure 2 due to their significant offsets from the main instrumentation grid. The location numbers for each shot location increase in order with the positive X direction for the SOS and SON locations, and with the positive Y direction for the SOW locations. Shots inside (SI) of the instrumented area are referred to by SI and the horizontal line directly south of them. For instance, the first line of shot locations furthest to the south and inside the instrumented area (refer to Figure 2) is named SIA. Similar to the outside shot locations, the inside shot location numbers along any given line increase with the positive X direction. For example, to aid the reader with orientation, shot locations SOW01, SOS01, SIL07, SIL12, and others are labelled in Figure 2. While the inside shot locations are organized on a regular 5-m x 5-m grid, T-Rex could not be used at all shot locations inside the instrumented area, either because of restricted maneuverability around trees or the presence of voids nearby that might collapse due to the weight of the truck. Thus, the SI locations illustrated in Figure 2 are clearly denoted as those where both T-Rex and impact sources were used and those where only impact sources were used (i.e., those where T-Rex could not be used). At each shot location where T-Rex was used, all three shaking modes were utilized: P-mode for vertical shaking, SL-mode for shear longitudinal shaking (i.e., in-line with the truck), and ST-mode for shear transverse shaking (i.e., cross-line to the truck). However, these three modes of shaking were not excited consecutively at each location. Instead, for each line of shots (e.g., SOS), a vibration mode was set (e.g., P-mode) and shots were performed along the entire line. Then, the shaking mode was switched (e.g., to SL-mode) and shot locations along the line were revisited using the updated mode. This was found to be more efficient than switching shaking modes at each shot location. Due to the significant number of shot locations and shaking modes used, and given the powerful nature of the source, no shot stacking was performed (i.e., only a single shot was collected for a given shaking mode at each shot location).

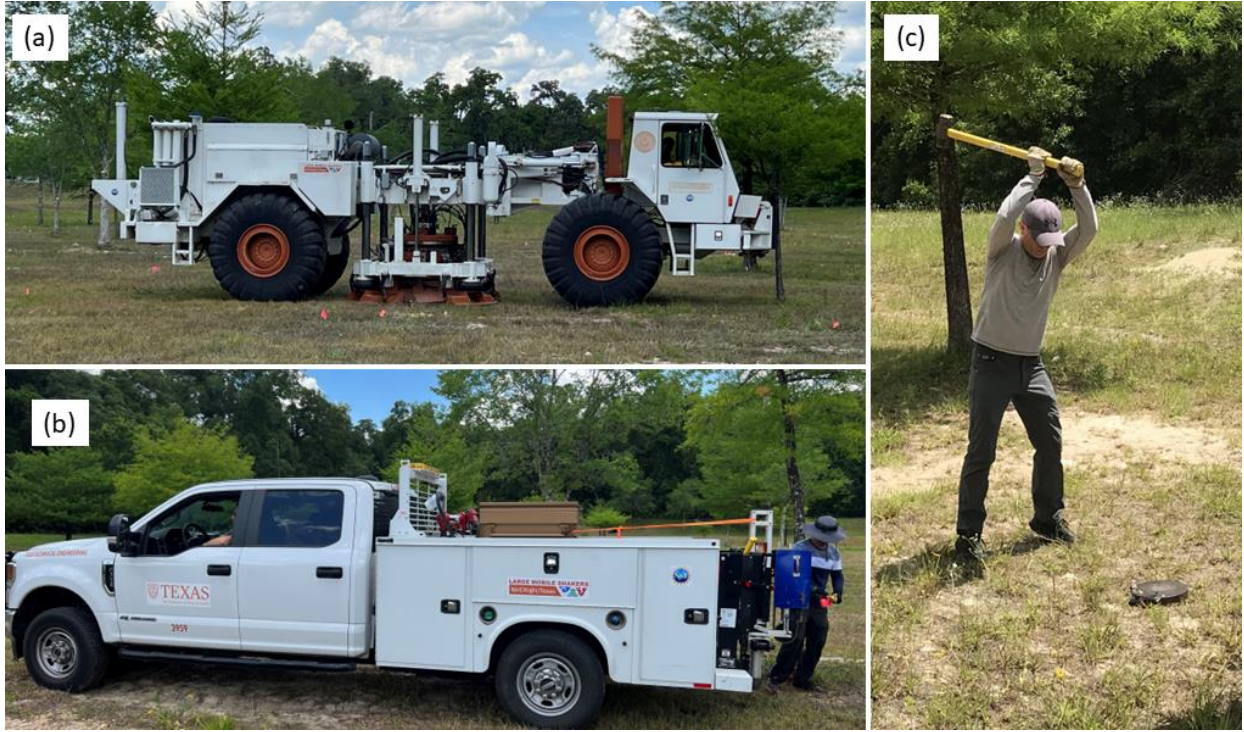

**Figure 5.** Pictures illustrating the various sources utilized in the experiment. Panel (a) shows the large, tri-axial vibroseis shaker truck, T-Rex, from the NHERI@UTexas experimental facility, panel (b) shows the PEG-40kg impact source mounted on a Ford F-350 pickup truck, and panel (c) shows an eight-pound sledgehammer.

The impact sources consisted of a 40-kg propelled energy generator (PEG-40kg), also known as an accelerated weight drop, manufactured by R.T. Clark Companies (Figure 5b), and an eight-pound sledgehammer (Figure 5c). The PEG-40kg is a portable source that generates seismic energy when a hammer mass is propelled downward by an elastomer band onto an impact plate. The hammer drop mass and height are 40 kg and ~40 cm, respectively, generally producing an impact frequency range of 10-250 Hz. There are two modes of operation for the PEG-40kg: single-cycle and continuous-cycle. For this experiment, the single-cycle mode was used. To ease mobility, the PEG-40kg was mounted on a Ford F-350 pickup truck, as shown in Figure 5b. The impact sources were only used to excite the ground inside the instrumented area, as shown in Figure 2. Since the F-350 pickup truck was significantly lighter and smaller than T-Rex, it was able to reach more shot locations. In the rare cases when the F-350 was unable to reach a shot location, an eight-pound sledgehammer was utilized instead of the PEG-40kg, ensuring that all inside shot locations in the 5-m x 5-m grid were covered by one of the impact sources.

**Passive-wavefield monitoring**

The DAS passive-wavefield data consisted of approximately four hours of ambient noise recordings on May 15$^{th}$ between 16:58 to 21:04 Universal Time Coordinated (UTC). During this period, there were two instances of rainfall which took place from 17:20 to 17:35 UTC and 20:54 to 20:56 UTC. Additionally, there was notable interference (i.e., noise) from an electric power generator on channels 30 to 66 and 1720 to 1742 during this time.

Every day from May 12$^{th}$ to May 16$^{th}$ the nodal stations were left to record ambient noise between 23:00 UTC, when work at the site was completed, to 11:00 UTC, when work resumed the next day. This resulted in a total of 48 hours of ambient noise data distributed over four, 12-hour time blocks gathered over a period of four days.

**Data acquisition**

*DAS*

An OptaSense ODH4+ IU was used in this experiment. The IU was borrowed from the NHERI@UTexas experimental facility (Stokoe et al., 2020) and was configured with a 2.04 m gauge length and 1.02 m channel separation, the minimum allowed by the OptaSense ODH4+. The gauge length refers to the average straight-line distance between the consecutive origins of the Rayleigh backscatter. Hence, the measurements of vibrations at each sampling location (channel separation) represent the average over the 2.04-m gauge length. The IU configuration and the cable length resulted in a total of 1920 DAS channels. In DAS it is desirable to set the laser pulse repetition (ping rate) as rapid as possible for a given cable length to increase signal-to-noise ratio and to mitigate phase demodulation errors. Phase demodulation errors arise when the strain change rate exceeds the DAS system's capability to detect, similar to how amplitudes that are too high for traditional seismographs result in clipping (Hubbard et al., 2022). This sometimes happens when a large source is activated near the fiber optic cable. To mitigate such errors, the ping rate can be set much higher than the desired time resolution/sampling rate, and the signal can be subsequently decimated to lower sampling rates relevant to the study being conducted in order to save memory space. On Friday, May 13$^{th}$, during the first day of active wavefield data acquisition, the OptaSense ODH4+ IU was configured with a (ping rate) of 50 kHz, and the DAS data was decimated to 10 kHz in real-time. On subsequent days, the ping rate was reduced to 20 kHz, and the data was decimated to 1 kHz. This lowering of the ping rate was necessitated by unanticipated problems with the laser pulse balancing on the ODH4+. Thus, the data collected on Friday, May 13$^{th}$ is of slightly higher quality because a 50 kHz ping rate was used. This should not be interpreted to mean that the data collected on other days is of poor quality due to it being acquired with a 20 kHz ping rate. Indeed, a 20 kHz ping rate is still more than adequate for interrogating a 2-km long cable. It's noteworthy that, on Friday, May 13$^{th}$, T-Rex executed all shots along the SOS and SOW lines.

The OptaSense ODH4+ IU outputs the raw data in consecutive one-minute H5 files that encompass all the DAS channels. Each channel contains the digitized output from the light phase difference over a gauge length, presented in radians relative to the average wavelength of the source light. Specifically, radians scaled by a factor of $2\pi/2^{16}$. H5 files utilize the hierarchical data format (HDF) to store vast amounts of data in the form of multidimensional arrays. At the end of the experiment, 538 GB of active and passive wavefield raw DAS data was collected and organized into three parent folders according to the source/excitation type: T-Rex, Impact sources, and Ambient noise.

*Nodal stations*

For this experiment, the SmartSolo nodal stations were configured with a sampling rate of 250 Hz. Every SmartSolo nodal station generated a folder, titled after its serial number, that housed three miniSEED files with amplitude units of counts, each file being associated with one of the station's three channels. These files contained the recording of the full duration of the experiment, from activation to deactivation of the stations. At the end of the experiment, the SmartSolo nodal stations had collected 48.6 GB of active-source and passive-wavefield data. The MiniSEED format, a commonly used subset of the Standard for the Exchange of Earthquake Data (SEED) format, was selected to facilitate data sharing and analysis. This format is widely recognized in the geophysical community, and there are open-source libraries like obspy and libmseed that provide support for reading and converting MiniSEED files to other formats. Thus, the raw data provided in the dataset comprises 144 folders, each corresponding to one of the 144 deployed nodal stations in the experiment.

**Supporting metadata documents**

The field handwritten notes were carefully reviewed, converted to digital format, and synthesized with other electronic notes and files to describe the dataset. All of this metadata has been organized in the "Supporting documents" folder. The information on the SmartSolo nodal stations, including serial numbers, station names, location codes, and both local and global coordinates, can be found in the "Nodal stations information" Excel file provided with the archived electronic dataset. The frequency instrument response curves for the vertical and horizontal geophones in the nodal stations are provided in the "DT-Solo5Hz(HP305) Frequency Response" Excel file. The "Cable information" Excel file contains the DAS channel indices, local and global coordinates for each corner along the fiber optic cable, as well as the cable's start and end points. The "T-Rex and impact sources information" Excel file provides information on all shots, whether from T-Rex or impact sources, including local and global coordinates for each shot, the shot location code, and the shot start time. Local and global coordinates for all the observed surface voids and trees in the vicinity of the instrumented area are provided in the "Voids and trees locations" Excel file. While these documents are best viewed using Excel due to the presence of illustrative images and multiple tabs for organizing information, alternative open-source CSV files are also provided, albeit without the images and advanced organization of the Excel files. A comprehensive and interactive PDF layout schematic (similar to what is shown in Figure 2, but with more information) has also been added to the supporting documents to enhance visual understanding of the experimental configuration. This multi-layered layout includes a representation of the fiber optic cable, nodal stations, and shot locations, with the assigned DAS channel for each cable corner and the experiment dimensions, voids, and trees locations all clearly marked.

**Processed data**

The electronic dataset publicly accessible through DesignSafe and documented herein contains both the raw and processed data from the DAS and SmartSolo sensing systems. While the raw data may be of interest to some readers, parsing and interpreting it in preparation for subsequent analyses can be a time-consuming task. Thus, we anticipate that the processed data will be of greater interest to most readers. Nonetheless, this paper and the accompanying electronic dataset provide all of the necessary information to completely re-interpret the raw data if desired. Figure 6 depicts the folder structure of the "Processed data" folder, showcasing two parent folders, "DAS" and "SmartSolo." Within these parent folders, the data is organized, based on the specific type of wavefield source, into three folders, namely: "T-Rex," "Impact sources," and "Ambient noise." Figure 6 also indicates the number of subfolders and files in each folder, along with the naming convention used for the files. The following subsections outline the processing steps used for the DAS and the SmartSolo nodal stations data and provide further details about the files stored in each folder and their naming convention.

*DAS*

As mentioned above, the DAS parent folder contains folders that correspond to data collected from T-Rex, Impact sources, and Ambient noise. For each T-Rex shot, the following processing steps were followed: (1) The shot start time was retrieved from the T-Rex electronics trigger file, where the GPS times for all shots were recorded. The times for the T-Rex shots automatically included a one-second pre-trigger delay. As noted above, these times are also provided in the "T-Rex and impact sources information" Excel file. (2) A 15-second window encompassing the shot was extracted from the one-minute-long DAS H5 files, which included a one-second pre-trigger delay, 12 seconds of T-Rex shaking, and two seconds of listen time post-T-Rex shaking to capture the waves reaching the array extremities. The 15-second windows were checked for dropped samples due to IU digitization and data storage errors in any of the 1920 DAS channels. Two of the 780 T-Rex shots had dropped samples on one or more channels. Specifically, Shots Z_SID11 and Z_SIF07 failed this check and were excluded from the processed dataset. (3) The raw data was scaled by $2\pi/2^{16}$ to convert it into phase change measurements in radians. (4) A 3-Hz high-pass filter was applied to remove low-frequency artifacts from laser drift and static strains caused by shaking close to the fiber optic cable, as recommended by Hubbard et al. (2022). (5) The DAS waveforms from the SOS and SOW shots were decimated from 10 kHz to 1 kHz to align with the rest of the collected data. (6) Phase data was converted to strain using Equation 1 (Hubbard et al. 2022):

$$\varepsilon_{xx} = \frac{\lambda \, d\phi}{4\pi n g \xi} \qquad (1)$$

where $\lambda$ is the average laser wavelength of the DAS system in a vacuum, equal to 1550 nm; $d\phi$ is the phase change measured by the DAS in units of radians; n is the group refractive index of the sensing fiber, approximately 1.47; $\xi$ is the photoelastic scaling factor for longitudinal strain in an isotropic medium, equal to 0.78; g is the gauge length, approximately 2.04 m; and $\varepsilon$ is the normal strain per single gauge length. (7) For each shot, the processed DAS data, along with the T-Rex pilot signal, base plate and mass accelerations, ground force in engineering units, and other shot-related information, such as local and global coordinates, shot time, and sampling rate, were organized into an "event" object and saved in an H5 file. The event object was created using a Python class, designed to efficiently organize the data collected during this experiment. To ensure that researchers with minimal programming proficiency can utilize the dataset, we have included Python tools in the Supporting documents folder for effortless querying and visualization of the H5 files (i.e., event objects). Each H5 file was given a unique label composed of T-Rex shaking direction, shot location, and shot timestamp. For example, file Z_SOS01_20220513154814.50 contains the data obtained when T-Rex was shaking vertically (i.e., in the P-mode Z-direction) at shot location SOS01 on 13 May 2022 at 15:48:14.5 UTC. The H5 files were organized in the T-Rex folder by shot location into two folders, "Inside shots" and "Outside shots", as illustrated in Figure 6. The Inside shots folder contains 535 H5 files, with a file for each T-Rex shaking direction for each of the 179-T-Rex inside shot locations (two shots were discarded as mentioned above). The Outside shots folder contains 243 H5 files, with a file for each T-Rex shaking direction for each of the 81 shot locations outside the array (refer to Figure 2). (8) All the DAS channels for each shot were cross-correlated with the T-Rex pilot signal and stored in H5 files, categorized into two folders: "Inside shots" and "Outside shots" based on the shot location. These folders contain the same number of files as the uncorrelated shots folders noted above. Additionally, the naming convention for the cross-correlated H5 files follow the same format as the uncorrelated shots, but with "CC" added to indicate "cross-correlated" (refer to Figure 6). For example, for the file described above, the cross-correlated data is stored in a file named Z_SOS01_CC_20220513154814.50, where CC distinguishes it from the uncorrelated records.

The same processing steps were followed for the impact sources, with a few notable exceptions. The shot start times were extracted from the field datasheet and verified through inspection of the closest nodal station waveforms. The impact arrival time at the nearest station was manually selected as the first instance of energy surpassing the noise floor. These times are provided in the "T-Rex and impact sources information" Excel file. The impact time was used to trim a three-second time window from the raw, one-minute H5 files, which included a one-second pre-impact portion and a two-second post-impact portion. The header of each processed H5 file for an impact source indicates if the shot was generated by an accelerated weight drop (i.e., the PEG-40kg) or the 8-lb hammer. The processed impact shot files were named using the same convention as noted above, but started with an "I" to denote "Impact" instead of an "X," "Y," or "Z" used to describe the T-Rex shaking orientation. These files are in a folder named "Inside shots" within the Impact sources folder to indicate that all the impact sources were excited inside the instrumented area. The Inside shots folder contains 286 H5 files corresponding to the 286 impact-source shot locations (refer to Figure 2).

The same processing steps followed for the active-source data were also followed for the four hours of ambient noise recorded by the DAS. However, the files were maintained as one-minute-long segments and labeled with the prefix "N" to distinguish them as noise recordings. These files are stored in the Ambient noise folder, which contains 246, one-minute-long H5 files that correspond to approximately four hours of ambient noise DAS recordings.

*Nodal stations*

The SmartSolo nodal station data processing was limited to extracting the shot time windows (15-second windows for T-Rex shots and 3-second windows for impact shots), merging the three individual components into a single miniSEED file, editing the header file information, and arranging them in a user-friendly format.

For T-Rex shots, the T-Rex trigger file was utilized to obtain the shot times, which were then used to trim a 15-second time window from all three components of the nodal stations for each shot. This included a one-second pre-trigger delay, a 12-second T-Rex shaking duration, and a 2-second listen segment post-T-Rex shaking. The T-Rex shot data collected from the nodal stations follows a comparable folder structure as that of the DAS, featuring two principal folders for "Inside shots" and "Outside shots". Nonetheless, each of these two folders is additionally subdivided into three folders based on the T-Rex shaking direction, namely "X Shaking", "Y Shaking", and "Z Shaking," corresponding to the local coordinate system. In each shaking direction folder for "Inside shots," individual folders are present for the 179-T-Rex inside shot locations. These folders encompass 145 files, out of which 144 correspond to the 144 nodal stations, while the remaining file comprises the T-Rex pilot signal, base plate and mass accelerations, and ground force in engineering units. Likewise, for the "Outside shots," the shaking directions folder comprise individual folders for the 81 outside shot locations, each of which contains 145 files. The miniSEED files for each nodal station include three components: DHN, DHE, and DHZ, following the naming convention recommended by FDSN (2012). The "D" in the name refers to the use of a 250 Hz sampling rate, "H" indicates the use of a high gain seismometer, and "N", "E", and "Z" indicate the geophone orientation (north, east, or vertical). The miniSEED header file holds important information, such as the record's sampling rate, start and end times, and the station location. The miniSEED files related to T-Rex shaking were named according to the format presented in Figure 6, including the T-Rex shaking direction, shot location, UTC date and time, station location, and the designation "3c" to indicate that the file includes the station's three perpendicular geophone records. For example, the file "X_SIA01_20220514143604.85_A01_3C.miniseed" is a 15-second miniSEED recording of T-Rex shaking in the X direction at shot location SIA01 on 14 May 2022 at 14:36:04.85 UTC captured by a nodal station located at A01. The miniSEED files containing the T-Rex source information follow the same naming convention, except they end with "Source.miniseed" instead of the station location name and "3c.miniseed".

The miniSEED files corresponding to the impact sources have a 3-second duration, with 1 second before the impact and 2 seconds following it. The SmartSolo data from the impact sources are organized in a folder named "Inside shots" within the Impact sources folder, following a structure similar to that of the DAS folder. Nevertheless, in contrast to the DAS folder structure, each of the 286 impact-source shots is stored in a separate folder. The impact source signature was not recorded. As a result, each folder for a given impact shot location contains 144 miniSEED files (one for each nodal station). The file naming convention is similar to that previously described for T-Rex shots, with the only difference being that the files begin with the letter "I" instead of the T-Rex shaking direction, as shown in Figure 6.

As noted above, the nodal stations were used to record 48 hours of ambient noise over a period of four days, in 12-hour increments from 23:00 to 11:00 UTC. The recorded data is stored in the Ambient noise folder, which is subdivided into four folders, one for each day during which the ambient noise was recorded. Each of these four folders comprises 1728 miniSEED files, consisting of 12 one-hour miniSEED files for each of the 144 nodal stations. These files are distinguished by the prefix "N" at the beginning, as depicted in Figure 6.

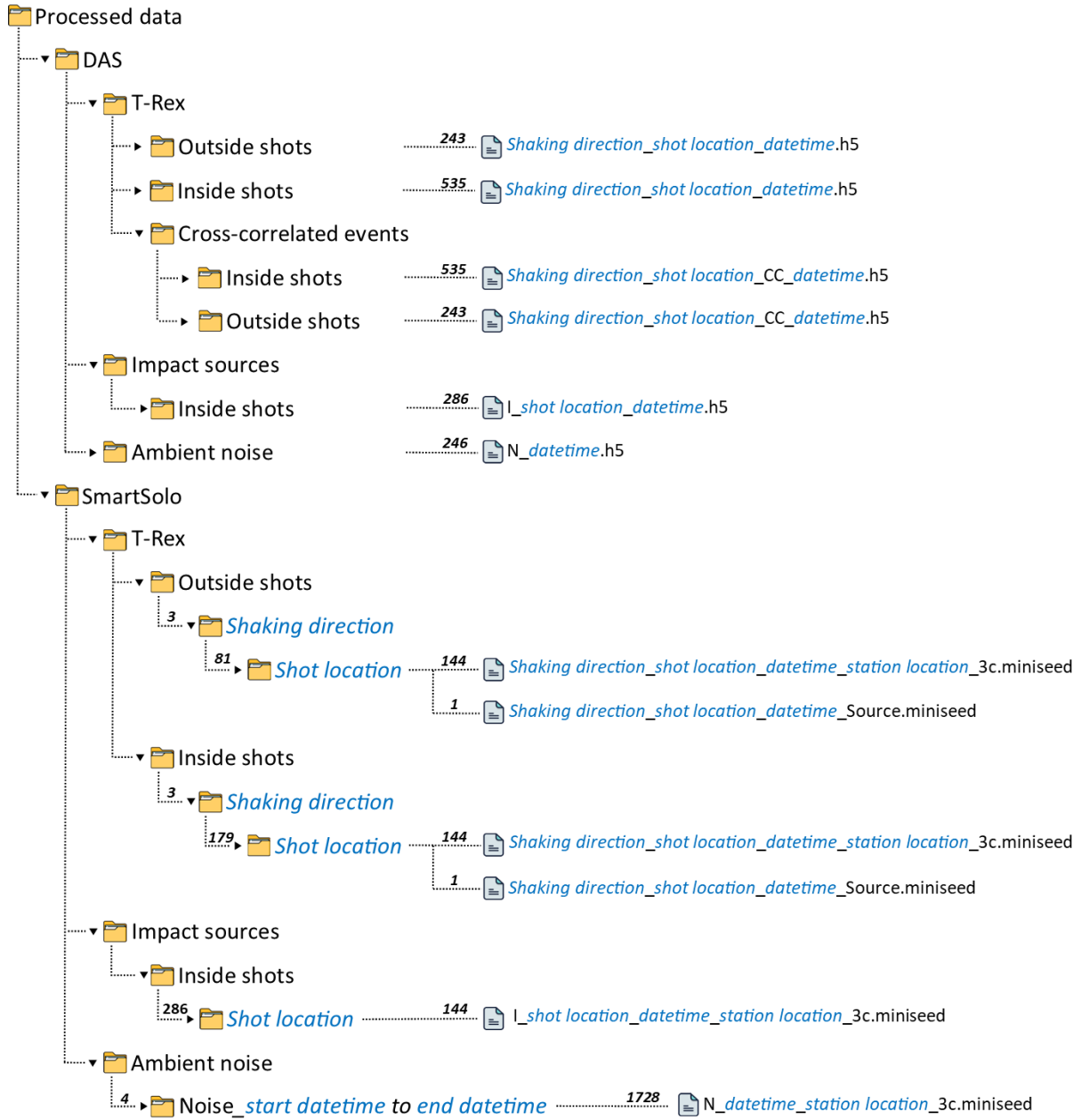

**Figure 6.** Schematic of the hierarchical folder structure for the processed data, indicating the number of subfolders and files within each folder. The blue, italicized text represents dynamic content, such as the T-Rex "shaking direction", which can be "X Shaking", "Y Shaking", or "Z Shaking", or the "shot location", which can be any of the numerous outside or inside shot locations, such as "SOS01". All dates and times are in Universal Time Coordinated (UTC).

**Potential Dataset Use Cases**

This section presents raw and pre-processed data examples from the archived, open-access dataset, with the aim of inspiring potential use cases for those interested in the dataset. The high-quality active-source data can be used for seismic migration, -refraction tomography, surface wave inversion, full-waveform inversion (FWI), and other imaging techniques. Similarly, the passive-wavefield data can be employed in techniques like horizontal-to-vertical spectral ratio (HVSR), microtremor array measurements (MAM), ambient noise tomography, and more. While no imaging results are presented in this section, we showcase that the waveforms and dispersion data extracted from the dataset are of high quality and can be used in such active-source and passive-wavefield imaging techniques. Additionally, we provide one or two example papers for each potential use case to aid readers in exploring further details.

*One-, two- and three-dimensional imaging*

Previous research conducted at this site by Tran and Hiltunen (2011), Tran et al. (2013), and (2020) revealed a high degree of spatial variability in its subsurface. This variability, coupled with the existence of karstic voids, suggests that employing 2D and 3D imaging techniques would be more suitable than 1D methods in effectively characterizing the subsurface. These imaging techniques can capitalize on the high spatial sensing resolution provided by the DAS, as well as the 3D sensitivity of the nodal stations, despite being more sparsely spaced in comparison to the DAS. Successful attempts have been reported in the literature in imaging the subsurface using active source 2D FWI (e.g., Wang et al., 2019; Tran et al., 2013) and 3D FWI (e.g., Fathi et al., 2016, Smith et al., 2019; Tran et al., 2020) using geophone data, and recently 2D FWI using DAS data (e.g., Yust et al., 2023). The efficacy of active-source imaging techniques critically depends on the wavefield generated by the source being sufficiently strong throughout the entire spatial extent of the instrumented area. Thus, rather than demonstrating specific imaging strategies/results, we instead focus on illustrating the quality of the collected waveforms and some potential ideas for taking advantage of the multi-direction sensing and multi-directional shaking on such a dense grid.

Figure 7 shows the waveforms generated by T-Rex shot X_SIL07 (refer to Figure 2) and recorded by the DAS channels and the DHE component of the nodal stations at the four furthest corners of the instrumented area (i.e., DAS channels 1787, 1733, 277, and 222, and nodal stations A01, A12, W01, and W12), which are labelled in Figure 2. These plots demonstrate that the wavefields generated by T-Rex were clearly sensed by both the DAS and nodal stations. Further evidence of this can be seen by examining a waterfall plot of the waveforms recorded by the longest DAS line (i.e., line 104) for T-Rex shot Y_SIL12, as shown in Figure 8. This waterfall plot presents the cross-correlated waveforms captured by DAS channels 31 through 183 (refer to Figure 2), with each trace normalized by its absolute maximum amplitude. Figure 8 reveals that clear waveforms were sensed by the entire DAS line. Furthermore, disturbances to the classical linear arrival-time moveout patterns can be observed in the spatially-dense DAS waveforms, potentially resulting from local heterogeneities and spatial variability at the site. The waveforms shown in Figures 7 and 8 are typical of the quality contained in this extensive dataset. As such, the waveforms should be more than adequate for performing various 1D, 2D and/or 3D imaging studies.

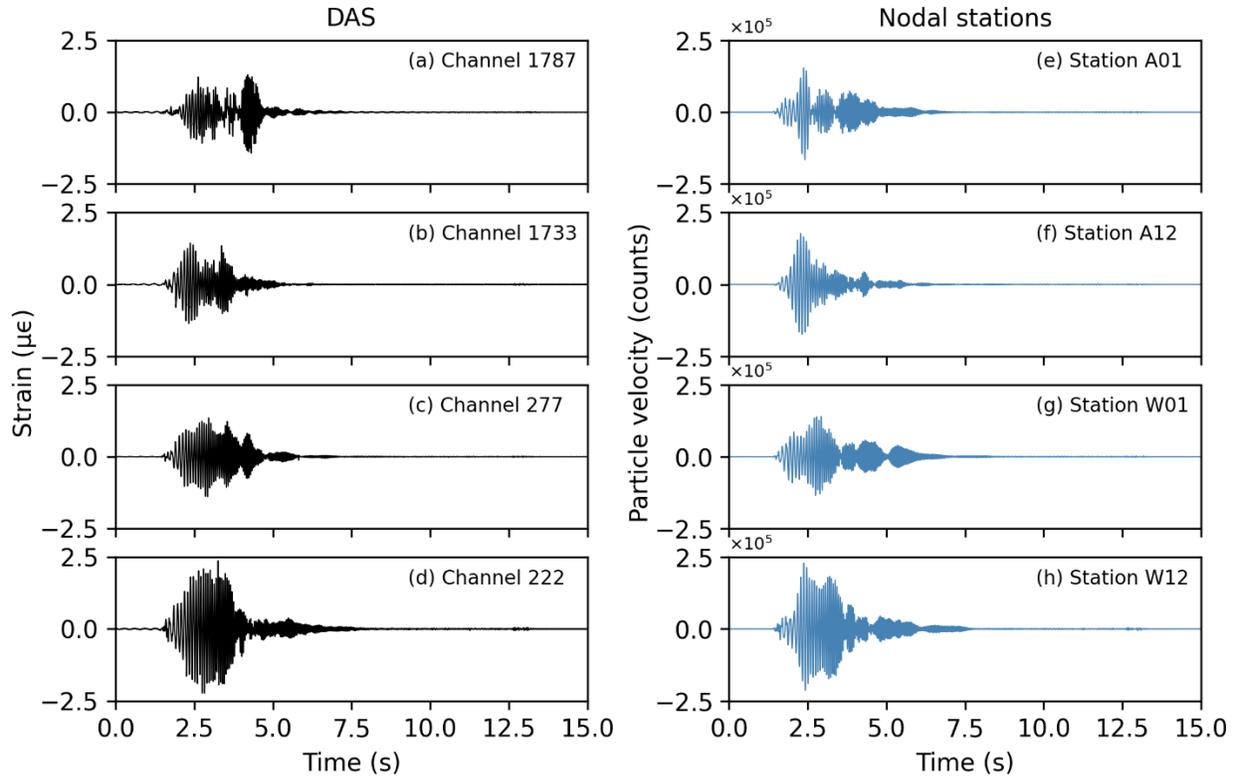

**Figure 7.** The waveforms generated by T-Rex shot X_SIL07 (refer to Figure 2), as captured by both the DAS channels and the DHE component of nodal stations positioned at the four furthest corners of the instrumented area (i.e., channels 1787, 1733, 277, and 222, and nodal stations A01, A12, W01, and W12).

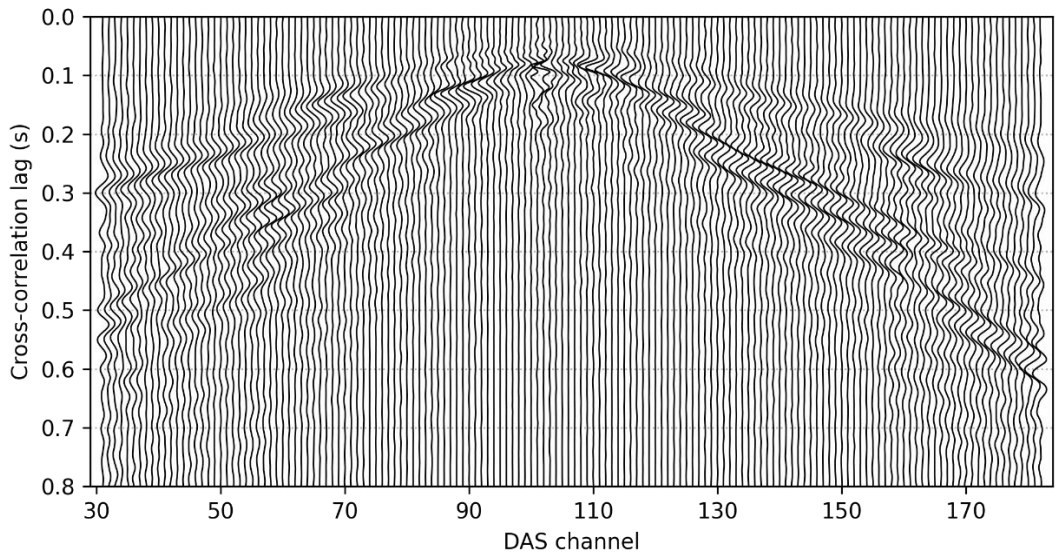

**Figure 8.** Waveforms recorded by DAS channels 31-183 along Line 104 (refer to Figure 2) after cross correlating with the T-Rex pilot signal for T-Rex shot Y_SIL12, and normalizing each waveform by its absolute maximum value.

The numerous shot locations and spatially-dense DAS channels and nodal stations provide ample opportunities to visualize wave propagation patterns in an attempt to locate anomalies prior to more rigorous processing, such as that used for psedu-2D MASW or 2D/3D FWI. One way to visualize the data is by observing the wave propagation as a function of time across an entire sensing array (i.e., either the DAS or the nodal stations) initiated by a T-Rex shot at one location. Alternatively, the wavefield recorded by a section of the sensing array, like a DAS line, generated by T-Rex shaking at multiple shot locations can be used to create wavefield animations. Figure 9 illustrates an example of the latter, where the cross-correlated waveforms generated by shots Z_SOW26, Z_SOW31, and Z_SOW37 and recorded by DAS channels 1810 through 1890 (refer to Figure 2) are shown, along with the location of a known surface void. In these waveforms, the peaks and troughs are filled with red and blue shading, respectively, to help better visualize disruptions to the wave polarities and wave propagation directions. Some of the backscatter events at the location of a known surface void can be observed in Figures 9a and 9c for shots Z_SOW26 and Z_SOW37, respectively, as highlighted by the dashed blue and red lines. Additionally, there is a noticeable increase in the amplitude of the late-arriving waves at the void location, as indicated by the intensity of the color and circumscribed by the dashed ellipses in Figures 9a and 9c. These observations are consistent with the outcomes of a synthetic study conducted by (Rector et al., 2015) that investigated a wavefield produced by an active source and measured by geophones situated on the surface directly above underground voids. Moreover, significant backscatter is evident between channels 1830 and 1840 for shots Z_SOW26 and Z_SOW31 (Figures 9a and 9b), which may indicate the presence of an underground anomaly, although no surface voids were observed at that location. The low signal strength in this zone could be caused by the presence of an anomaly and/or poor cable coupling. The waveforms from this zone need to be examined from other shot locations in order to conclusively determine the cause(s).

Regarding the potential to use the dataset for 1D and/or pseudo-2D surface wave imaging; Figure 10 shows a selection of surface wave dispersion images that can be derived from various permutations of T-Rex shaking directions at just one-shot location, namely SOW40, and utilizing a single line, line Q, of DAS channels and nodal stations. The locations of shot SOW40 and line Q are shown relative to the entire experiment in the schematic map presented in Figure 10a. The dispersion images displayed in Figure 10 were generated using the frequency-domain beamformer (FDBF) technique with cylindrical-wave steering, square-root-distance weighting (Zywicki and Rix, 2005), and frequency dependent normalization, as implemented in the open-source Python package swprocess (Vantassel, 2022). Figures 10b and 10c display dispersion images derived from channels DHN (crossline) and DHZ (vertical) in nodal stations Q01 through Q12 for shots Y_SOW40 and Z_SOW40, respectively, conforming to the customary multi-channel analysis of surface waves (MASW) configuration used for processing Love and Rayleigh waves with nodal stations, respectively. The peak power points at each frequency are indicated by white dots. Clear fundamental mode Love ($L_0$) and Rayleigh ($R_0$) wave trends are visible in the dispersion images in Figures 10b and 10c, respectively, along with some potentially $1^{st}$-higher mode Rayleigh wave trends (R1?) trends.

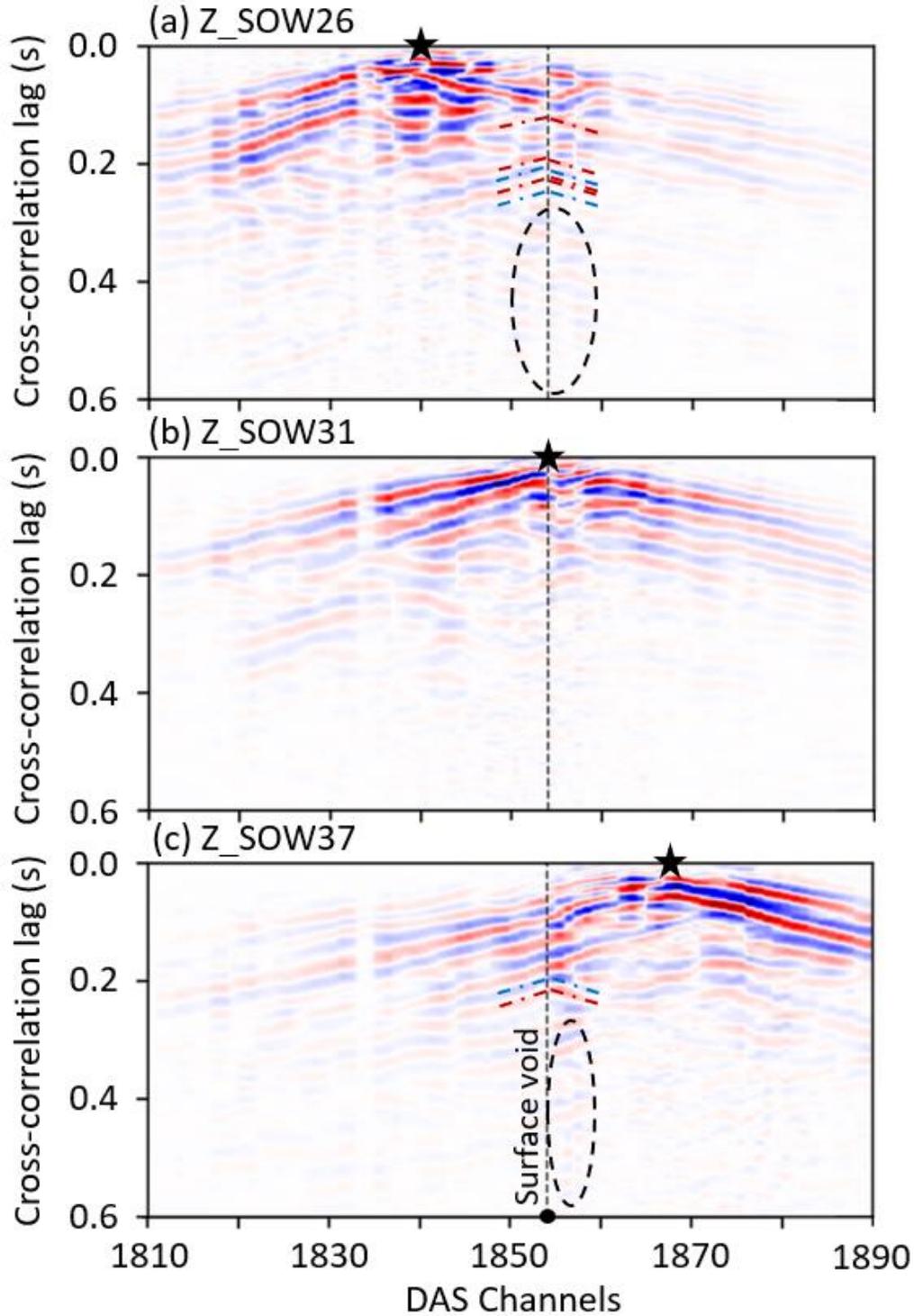

**Figure 9.** Waveforms recorded by DAS channels 1810 through 1880 for shots Z_SOW26, Z_SOW31, and Z_SWO37 (refer to Figure 2) in Panels (a), (b), and (c), respectively. A surface void location is indicated by a black dashed line in all panels, with backscatter evident at its location, highlighted by dashed blue and red lines. The dashed ellipses in panels (a) and (c) circumscribe the relatively higher amplitude of the late-arriving waves at the void location compared to other locations in the wavefield at similar time lags, as indicated by the intensity of the color.

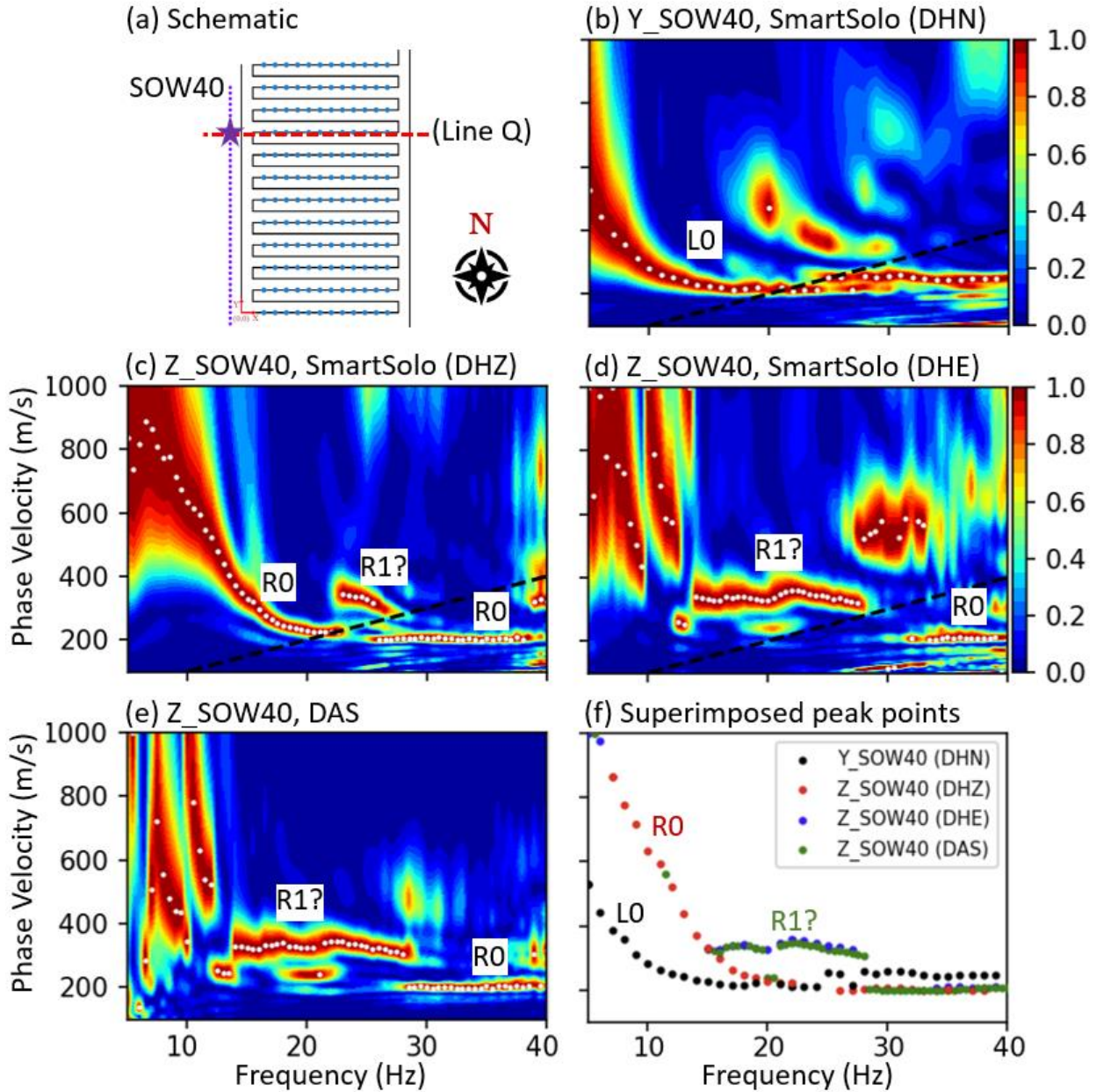

**Figure 10.** Dispersion images obtained from DAS and nodal stations along line Q due to T-Rex shaking at location SOW40. Panel (a) presents a schematic map highlighting the locations of shot location SOW40 and line Q. Panels (b), (c), and (d) showcase dispersion images from nodal stations Q01 through Q12 for shots Y_SOW40, Z_SOW40, and Z_SOW40, respectively, derived from DHN, DHZ, and DHE components, respectively. Panel (e) displays

While Figure 10c illustrates the more common approach to calculating Rayleigh wave dispersion images (i.e., using the vertical components to record the wavefield from a vertical source), theoretically, Rayleigh wave dispersion data could be obtained from a vertical excitation using either the vertical particle motion or the horizontal inline particle motion (Vantassel et al., 2022). Figures 10d and 10e present dispersion images obtained from the DHE (inline) channels in nodal stations Q01 through Q12 and DAS channels 629 through 693, respectively, for shot Z_SOW40. This configuration of vertical shaking and horizontal inline DAS channels is typically employed as a standard MASW arrangement for processing Rayleigh waves with DAS, however, the use of inline geophones for Rayleigh waves is not common practice. Nonetheless, for comparative purposes with the DAS Rayleigh wave dispersion image, we include them in this study, as previously done by Vantassel et al. (2022). Despite the disparity in the number of Line Q DAS channels (65) and nodal stations (12) utilized in developing the dispersion images in panels 10d and 10e, respectively, the resemblance between the resulting dispersion images is evident, which agrees with the observations of Vantassel et al. (2022). Nonetheless, the DAS line offers a significant advantage in such a spatially variable site, as it enables 2D MASW-type processing (M. Yust et al., 2022), which is not feasible using the nodal stations due to the limited number of stations deployed at each line. Comparing the dispersion image in Figure 10c to the ones in Figures 10d and 10e, it seems that the vertical particle motion recorded by the SmartSolo DHZ components resolves more of the apparent $R_0$ trend. However, there is a benefit to using both the vertical and horizontal particle motions to obtain a better understanding of the Rayleigh wave propagation. For example, the benefits of integrating Rayleigh wave dispersion data from both vertical and horizontal particle motions becomes apparent when examining Figure 10f which compares the dispersion data obtained from combining the peak power trends from all four dispersion images in Figure 10b – 10e. This combined approach enables a more clear identification of fundamental and higher mode trends and clearly highlights the anticipated higher Love wave phase velocities at higher frequencies, as compared to Rayleigh waves (Soomro et al., 2016). In summary, the multi-direction shaking and multi-component sensing allows for more robust surface wave dispersion processing.

The dataset can also be utilized in machine learning imaging studies, which have been gaining significant interest in the last few years. For instance, recent studies have showcased the potential of applying convolutional neural networks (CNNs) to image the near surface (e.g., Abbas et al., 2023; Crocker et al., 2023; Vantassel et al., 2022). Vantassel et al. (2022) trained a CNN to take a wavefields inputs and generate 2D $V_S$ images of the near surface, while Abbas et al. (2023) developed a CNN that employs dispersion images as inputs to produce 2D $V_S$ near surface images that was validated on field data. The high-quality waveforms and their derived dispersion images shown in Figures 7 through 10 demonstrate the potential of using the dataset in such machine learning studies.

In the preceding paragraphs, potential use cases for the application of active-wavefield data in imaging have been presented. However, it should be noted that successful imaging of the subsurface using passive-wavefield data has also been documented in the literature. For instance, the 3-component (3C) noise data could be utilized in 2D microtremor array measurements (e.g., Wathelet et al., 2018) and 2D/3D ambient noise tomography (Wang et al., 2021, 2023). Additionally, the horizontal-to-vertical spectral ratio (HVSR) measurement technique has been demonstrated to provide valuable information about the subsurface, as discussed in the following section.

*Horizontal-to-Vertical Spectral Ratio (HVSR)*

HVSR can be used to infer the spatial variability of fundamental site period ($T_0$) at each 3C nodal station. The fundamental site period can reveal important information about the location/depth of strong impedance contrasts beneath each nodal station (Bard & SESAME Team, 2004). For example, stations with higher $T_0$ values (or lower fundamental frequency, $f_0$) are expected to have deeper impedance contrasts, while stations with lower $T_0$ values (or higher $f_0$) are expected to have more shallow impedance contrasts. The calculation of HVSR can be accomplished using the passive-wavefield data collected by the nodal stations. The HVSR was computed using only one hour of ambient noise recordings captured on the 14 May at 4:00 UTC (i.e., files N_20220514040000_*station location*_3C). The data was processed with the open-source Python package hvsrpy (Vantassel, 2021). The one-hour long recording for each station was divided into 30, 120-second-long time windows and the horizontal components were combined using the geometric-mean, as recommended by (Cox et al., 2020) Further, smoothing was performed using the filter proposed by (Konno & Ohmachi, 1998) with b=40. A color-mapped representation of the 144 stations' fundamental frequency from the HVSR median curve ($f_{0,mc}$) can be found in Figure 11, indicating a notable fluctuation in $f_{0,mc}$ throughout the site, with a range of values spanning from 4.19 to 7.03 Hz. HVSR amplitude with frequency plots for stations A01, M01, and W01 are also shown in Figure 11. Higher $f_{0,mc}$ values are indicative of shallower depths to limestone, while lower $f_{0,mc}$ values correspond to deeper depths. This lends additional support to the site's significant spatial variability, as concluded by Tran and Hiltunen (2011) and Tran et al., (2013) and (2020). The HVSR data could be processed in more rigorous ways to extract more qualitative estimates for the depth to bedrock (e.g., Bignardi et al., 2016; Hobiger et al., 2009; Scherbaum et al., 2003).

*DAS reception patterns*

Using one-dimensional strain measurements to detect stress waves can greatly impact the waves measured phase and magnitude, owing to their directional sensitivity. The sensitivity of a DAS array to stress waves depends on the angle at which the waves impinge on the cable and the ratio between the wavelength and the gauge length (Martin et al., 2021). It is therefore essential to study and consider this phenomenon when using DAS for active source stress wave measurements. Martin et al. (2021) developed a comprehensive analytical full waveform representation of pointwise and distributed strain-rate measurements for all kinds of planar surface and body waves. Similarly, Hubbard et al. (2022) developed numerical representations of DAS reception patterns for different source orientations and wavelength-to-gauge-length ratios. Figure 12a demonstrates their work for a wavelength-to-gauge-length ratio of five, depicting the horizontal strain measurements ($\varepsilon_{xx}$) resulting from X-direction Ricker wavelet excitation caused by a point force at the surface. The $\varepsilon_{xx}$ values shown in Figure 12a resemble those that a horizontally placed DAS cable oriented in the X direction would measure. In this representation, red indicates tension while blue represents compression. Notably, a distinct change in wavefield polarity is evident between the left and right sides of the shot location in Figure 12a, with a zone of zero sensitivity directly above and below the shot location (i.e., at 90 and 270 degrees from the zero X axis). The dataset documented herein offers a valuable resource for analyzing DAS reception patterns using real field data, thanks to its abundance of shot locations and shaking directions, as well as the utilization of a dense 2D DAS array. Figure 12b displays a snapshot of the waveforms recorded on DAS channels 218 through 1791 at 1.9 seconds into shot X_SIL07 (refer to Figure 2). In this figure, a clear polarity flip can be observed between the right and left of the shot location, particularly between the two wavefronts marked by the dotted black circles. To highlight this contrast, we inverted the polarity of all channels to the right of the shot location by multiplying their values by -1. Figure 12c shows the results from this reversal of the DAS polarity on the channels to the right of the source. After flipping the polarity of the channels to the right of the source location, the two dotted circles in Figure 12c mainly encompass a tension wave propagating away from the source, as inferred by the red cable color. By reproducing Hubbard et al.'s (2022) numerical simulations with real-field data, Figure 12b underscores the potential of the dataset for investigating DAS reception patterns and any potential effects of underground anomalies on them.

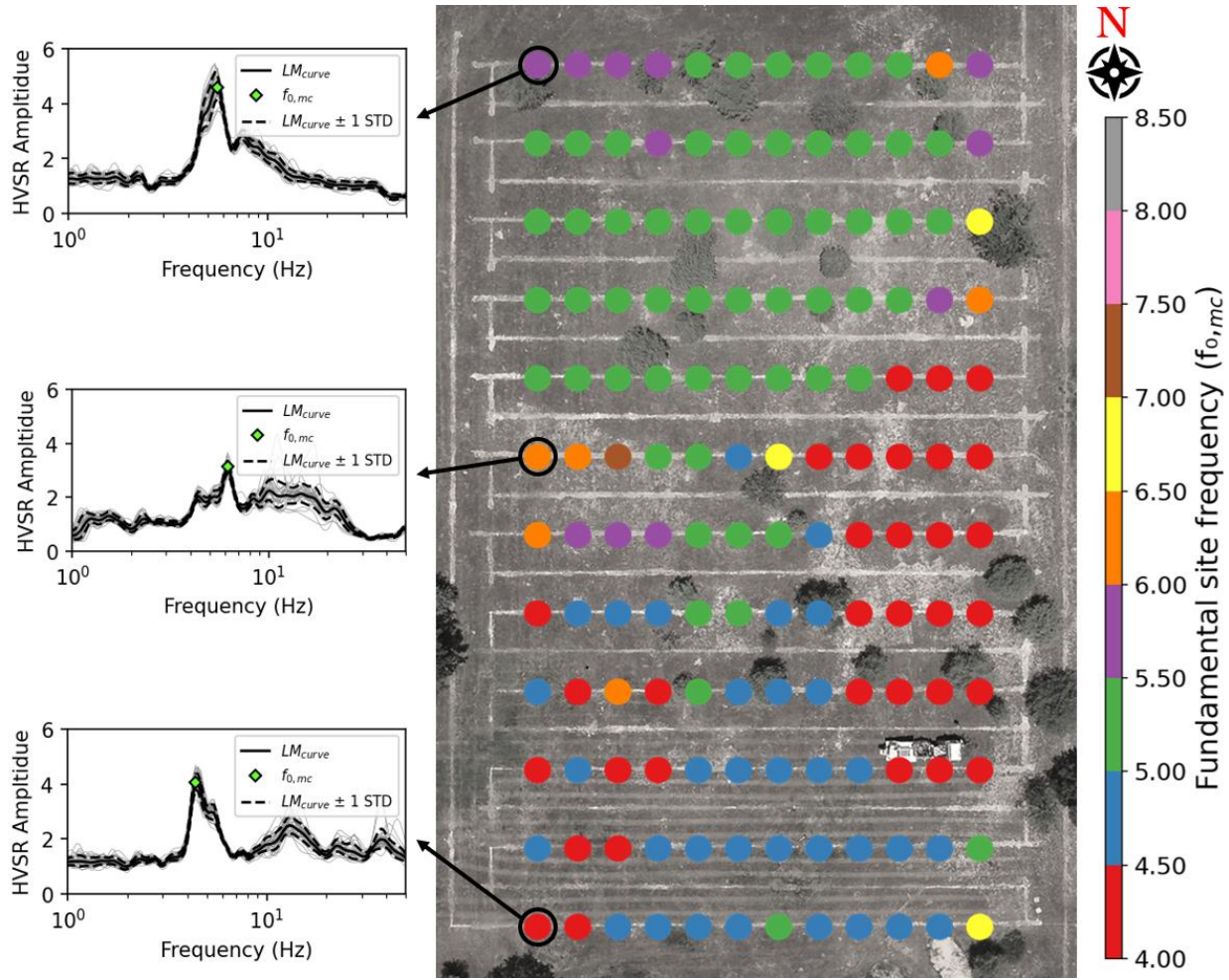

**Figure 11.** Spatial distribution of the fundamental site frequency ($f_0$), as determined by the peak of the lognormal median curve ($f_{0,mc}$), obtained from the Horizontal-to-vertical spectral ratio (HVSR) analysis of one hour of ambient noise data collected at each nodal station. Detailed HVSR plots are shown for selected nodal stations (i.e., A01, M01, and W01), depicting the HVSR calculations for each time window, the lognormal median curve ($LM_{curve}$), the ± 1 lognormal standard deviation (STD) curves, and the fundamental site frequency from the median curve ($f_{0,mc}$).

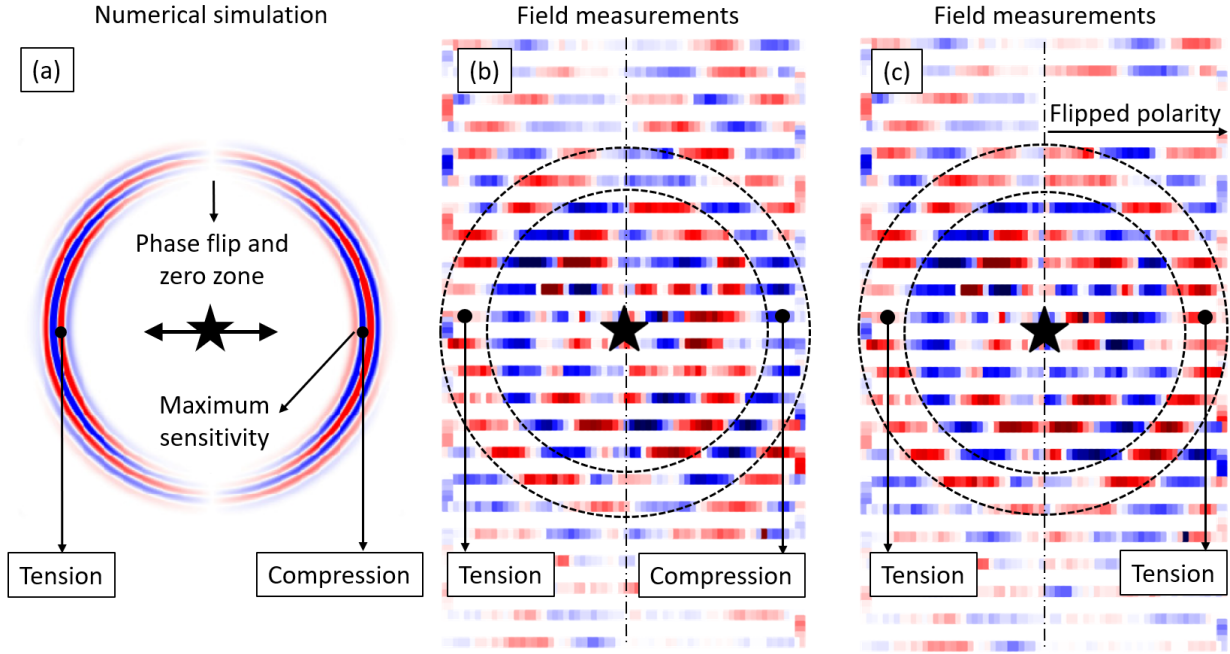

**Figure 12**. Reception patterns of DAS due to a horizontal excitation at the ground surface, as obtained from a numerical simulation in Panel (a) and field measurements in Panels (b) and (c). Panel (a) displays the numerical x-direction surface strain ($\varepsilon_{xx}$) results, with compression shown in blue and tension in red, obtained from an elastic half space excited by a Ricker wavelet in the X direction with wavelength-to-gauge length ratio ($\lambda/g$) of five, after Hubbard et al. (2022). Panel (b) displays the DAS field measurements, as detected by channels 218 through 1791, 1.9 seconds after the initiation of T-Rex-induced shaking in the X direction at location SIL07. Panel (b) also highlights a clear reversal of polarity between the left and right sides of the shot, which is best observed by examining the colors between the wavefronts indicated by dotted lines. To accentuate the distinction, Panel (c) replicates Panel (b) after the polarity of channels to the right of the shot location have been flipped.

**Conclusions**

This research paper outlines a comprehensive subsurface imaging experiment in Newberry, Florida using stress waves. The site is spatially variable and contains karstic surface and underground voids and anomalies, which have been documented in the literature and indicated through preliminary processing of the collected data. The sensing technologies used comprised a dense 2D array of 1920 DAS channels and a 12 x 12 grid of 144 SmartSolo 3C nodal stations, which covered an area of 155 m x 75 m and were used to record both active-source and passive-wavefield data. The active-source data was generated by a variety of vibrational and impact sources, namely: a powerful three-dimensional vibroseis shaker truck, a 40-kg propelled energy generator (PEG-40kg), and an 8-lb sledgehammer. The vibroseis shaker truck was used to vibrate the ground in the three directions at 260 locations inside and outside the instrumented area, while the impact sources were used at 268 locations inside the instrumented area. In addition to active source data, four hours of ambient noise were recorded using the DAS, while the nodal stations recorded 48 hours of ambient noise in four 12-hour increments over a period of four days. The waveforms obtained from the 1920 DAS channels for every active-source shot or passive-wavefield time block were extracted, processed, and stored in H5 files. These files can be easily visualized using a Python script incorporated with the open-access dataset. Additionally, the three-component data gathered from each SmartSolo nodal station were consolidated into a single miniSEED file, and the data from all 144 nodal stations obtained during each active-source shot or passive-wavefield time block were extracted and saved into a separate folder. To enable efficient retrieval of all necessary information, the dataset was systematically organized into three parent folders - raw data, processed data, and supporting documents - with a consistent naming convention employed for all files and folders. The raw and processed dataset, along with complete and detailed documentation of the experiment, have been archived and made publicly available on DesignSafe. We anticipate that this dataset will be a valuable resource for researchers developing techniques for void and anomaly detection using noninvasive, stress wave-based subsurface imaging techniques. We have provided examples of the data and potential use cases as a means to inspire present and future researchers who need a high-quality experimental dataset with known and unknown anomaly locations for testing imaging methods.

**Acknowledgements**

The authors would like to acknowledge Dr. FarnYuh Menq, Mr. Robert Kent, Mr. Andrew Valentine, Dr. Ruoyu Chen and Dr. Scott Wasman for assisting in the field data acquisition. This work was supported by the U.S. National Science Foundation grants CMMI-2120155 and CMMI-1930697. However, any opinions, findings, conclusions or recommendations expressed in this material are those of the authors and do not necessarily reflect the views of the National Science Foundation.